# Macroscopic Quantum Tunneling in Superconducting Junctions of $\beta$-Ag$_2$Se Topological Insulator Nanowire

*Jihwan Kim,[1,†] Bum-Kyu Kim,[2,†] Hong-Seok Kim,[2] Ahreum Hwang,[1] Bongsoo Kim,[1,\*] Yong-Joo Doh[2,\*]*

[1]Department of Chemistry, KAIST, Daejeon 34141, Korea

[2]Department of Physics and Photon Science, Gwangju Institute of Science and Technology (GIST), Gwangju 61005, Korea

**ABSTRACT**

We report on the fabrication and electrical transport properties of superconducting junctions made of $\beta$-Ag$_2$Se topological insulator (TI) nanowires in contact with Al superconducting electrodes. The temperature dependence of the critical current indicates that the superconducting junction belongs to a short and diffusive junction regime. As a characteristic feature of the narrow junction, the critical current decreases monotonously with increasing magnetic field. The stochastic distribution of the switching current exhibits the macroscopic quantum tunneling behavior, which is robust up to $T = 0.8$ K. Our observations indicate that the TI nanowire-based Josephson junctions can be a promising building block for the development of nanohybrid superconducting quantum bits.

Topological insulators (TIs) are bulk insulators with gapless metallic surface (or edge) states, which are topologically protected by time-reversal symmetry.[1] The topological surface states are known to have a spin-helical nature, meaning that the electron spin is aligned parallel to the surface and perpendicular to the translational momentum. As a result of the spin-momentum helical locking, the surface states in TIs are protected from electronic backscattering,[2] and they are thus expected to exhibit highly quantum-coherent charge and spin transport. The chiral nature of TIs is therefore promising for quantum information devices[3] and spintronics applications.[4]

Combined with conventional *s*-wave superconductors, TIs are expected to provide a useful platform for creating and manipulating the Majorana bound state,[5] which is essential for fault-tolerant quantum information processing.[6] Establishing highly-transparent contacts with superconducting electrodes on the TI surface enables the observation of supercurrent through the TI via the superconducting proximity effect.[7] To date, the Josephson supercurrent has been observed in various TIs such as $Bi_2Se_3$,[8-10] $Bi_2Te_3$,[11, 12] $Bi_{1.5}Sb_{0.5}Te_{1.7}Se_{1.3}$,[13] $Bi_xSb_{2-x}Se_3$,[14] strained HgTe,[15] and InAs/GaSb quantum wells.[16] Moreover, the peculiarity of topological superconductivity, the so-called 4π-periodic Josephson effect, has been manifested by the observations of anomalous microwave response[17-19] and abnormal oscillations of the critical current in TI-based superconducting quantum information device (SQUID).[20]

The switching event from a supercurrent branch to a dissipative quasiparticle branch in a Josephson junction (JJ) can be understood by the escaping motion of a fictitious phase particle confined in a tilted washboard-type potential well (see the inset of Fig. 3b and the Methods),[7] in which the phase is determined by a superconducting phase difference between two superconducting electrodes in the junction. The barrier height, $\Delta U$, is proportional to the maximum supercurrent, i.e., the critical current $I_C$ of the JJ. In a parabolic approximation of the potential well, there exist three different regimes of the escaping motion, such as macroscopic quantum tunneling[21, 22] (MQT), thermal activation[23] (TA), and phase diffusion[24, 25] (PD) processes, depending on thermal fluctuations and the Josephson coupling strength. In particular, observation of the MQT provides direct evidence for the quantum nature of the JJ, which is essential for developing superconducting quantum information devices.[26] In contrast to conventional tunnel-type JJs,[27] nano-hybrid JJs made of semiconducting nanostructures[28-30] can be used to introduce an additional control knob to tune the MQT behavior through the gate-tunable $I_C$, which has been demonstrated using graphene[31, 32] and InAs nanowire[33] (NW). In this work, we report the stochastic switching-current distributions in the MQT regime, obtained from $\beta$-Ag$_2$Se TI NW-based JJs. To the best of our knowledge, the MQT nature inherent in TI JJs is reported for the first time. Moreover, the crossover temperature to observe the MQT behavior reaches up to $T^*_{\mathrm{MQT}} = 0.8$ K, which is the highest among the nanohybrid JJs. Our work could pave the way for developing gate-tunable superconducting qubits[34, 35] made of TI NWs.

**RESULTS AND DISCUSSION**

Single crystalline $\beta$-Ag$_2$Se NWs were grown on a $c$-Al$_2$O$_3$ substrate by chemical vapor deposition (Figure 1a).[36] The NWs had clean surfaces and clear facets with diameters of 100 ~ 250 nm (Figure 1a inset). The high-resolution transmission electron microscopy (TEM) and selected-area electron diffraction (SAED) pattern reveal single crystallinity of the $\beta$-Ag$_2$Se NWs. The lattice space of 2.620 Å is well matched to the (003) plane of the orthorhombic $\beta$-Ag$_2$Se crystal structure (Figure 1c) and (113), (223), and (110) planes in the SAED pattern are also matched to orthorhombic $\beta$-Ag$_2$Se along the [1$\bar{1}$0] zone axis (Figure 1d). The X-ray diffraction pattern is also indexed to the $\beta$-Ag$_2$Se structure (see Figure S1).

A single $\beta$-Ag$_2$Se NW was transferred from a sapphire substrate onto a highly doped $n$-type silicon substrate covered by a 300-nm-thick oxide layer. Conventional electron-beam lithography and RF sputtering were used to form Ti (15 nm)/Al (150 nm) electrodes. The scanning electron microscope (SEM) image of a typical superconducting junction made of a $\beta$-Ag$_2$Se NW is displayed in the inset of Figure 2a, where the channel length and diameter are $L$ = 145 nm and $\phi$ = 175 nm, respectively. The electrical transport properties were measured using a $^3$He refrigerator (Cryogenic Ltd.) with a base temperature of $T$ = 0.3 K. The switching current distribution over 4,000 repetitions was obtained using a triangle-wave-shaped current with a ramping rate $dI/dt$ = 238 µA/s and a threshold voltage of $V_{th}$ = 30 µV. The existence of topologically non-trivial surface states in the $\beta$-Ag$_2$Se NW has been confirmed by measuring electronic transport properties, including the weak antilocalization effect, Aharonov-Bohm oscillations, and Shubnikov-de Haas oscillations in our previous work.[36]

Temperature-dependent current–voltage ($I$–$V$) characteristic curves for device **D1** are displayed in Figure 2a, where the bias current was swept from negative to positive values. We note that the $I$–$V$ curve at the base temperature, $T$ = 0.3 K, exhibits a dissipationless branch up to the critical current $I_C$ ~ 29 µA, which corresponds to the critical current density of $J_C$ ~ 1.3 × 10$^5$ A/cm$^2$. Device **D2** also exhibits similar $I_C$ and $J_C$ values (see Fig. S2). Our observed $J_C$ is about seven times larger than the one from the shortest channel (~ 30 nm) device[37] of an InAs NW junction, and twice as large as that from the Nb-contacted InN NW junction.[38] To the best of our knowledge, the $I_C$ and $J_C$ values observed in this work are the highest values among the results obtained from the semiconductor-NW-based superconducting junctions to date.[28, 37-40] We also note that there occurs an abrupt voltage jump from supercurrent to resistive branches at $I_C$, while the reversed voltage drop is observed at the return current $I_R$. The hysteretic $I$–$V$ curves obtained at low temperatures can be caused by an effective capacitance in the nanohybrid superconducting junction[41] or Joule heating effect.[42] Here, the power density near $I_C$ amounts to $P \sim 4.8$ μW/μm$^3$ at $T = 0.3$ K, which is large compared to those in other reports.[42] Thus, we infer that the electron temperature increases abruptly just above $I_C$, and the elevated temperature drops to the bath temperature value below $I_R$, resulting in the hysteretic $I$–$V$ curves.

The temperature dependence of $I_C$ and $I_R$ is displayed in Figure 2b. We note that the $I_C(T)$ plot has a convex shape instead of an exponentially decaying one, which is typically observed in other nanohybrid superconducting junctions[40, 41, 43]. The behavior of the former indicates that our β-Ag$_2$Se JJ is within the short junction limit,[44] while the latter is consistent with the long and diffusive junction limit.[45] As the elastic mean free path of β-Ag$_2$Se NW,[36] $l_e \sim 21$ nm, is considerably shorter than the channel length $L$, we used a short and diffusive junction model to fit our $I_C(T)$ data (see the Methods). The calculation result (dashed line in Figure 2b) is in qualitative agreement with the experimental data. Here, the $I_C R_N$ product reaches about 770 μeV for **D1**, resulting in $eI_C R_N / \Delta_{Al} = 2.9$, where $R_N = 27$ Ω refers to the normal-state resistance of the junction and $\Delta_{Al} = 260$ μeV is the superconducting gap energy of Al, which is estimated from the superconducting transition temperature $T_C = 1.7$ K.

The application of a magnetic field $B$, perpendicular to the substrate, induces a progressive change of the $I$-$V$ curves (see Figure S3). A color plot of the dynamic resistance, $dV/dI$, in Figure 2c depicts the supercurrent region (dark blue) in contrast to the normal-state region (light blue), while the *dV/dI* peaks (dark red) occur at $I_C$. In particular, Figure 2d shows that $I_C(B)$ data exhibit a monotonically decreasing behavior with *B* in Figure 2d, instead of a periodic modulation with a period of $B_0 = \Phi_0/Ld \approx 13$ mT, where $\Phi_0 = h/2e$ is the magnetic-flux quantum. Such monotonous behavior of $I_C(B)$ data has also been observed in other NW-based JJs,[37-40, 46] which can be well understood within the narrow junction model.[47] When the width (or diameter of NW) of the junction is less than the magnetic length, $\xi_B = (\Phi_0/B_0)^{1/2}$, the application of a magnetic field can break the proximity-induced superconductivity in the NW. For device **D1**, $\xi_B = 160$ nm is close to $\phi = 175$ nm, which is in reasonable agreement with the condition of the model. The theoretical calculation (solid line) fits the $I_C(B)$ data very well (see the Methods), resulting in the Thouless energy of $E_{Th} = \hbar D/L^2 = 22$ μeV, where $D = v_F l_e/3$ is the diffusion coefficient and $v_F$ the Fermi velocity. Using[36] $D = 22$ cm$^2$/s, we obtain an effective junction length of $L^* = 250$ nm, which is considerably longer than *L*. The difference is attributed to an effective elongation of the junction into the superconducting electrodes by the amount of the superconducting coherence, $\xi = (\hbar D/2\Delta_{Al})^{1/2} = 52$ nm.

Figure 3a depicts a stochastic distribution of $I_C$ at $T = 0.3$ K after 4,000 iterations of current sweep with a voltage criterion of $V_{th} = 30$ μV. As the temperature is increased, the $I_C$ distribution varies distinctively, as can be seen in Figure 3b, i.e., there are very sharp distributions at higher temperatures near $T_C = 1.7$ K, much broadened ones at intermediate temperatures, and moderate and temperature-independent ones at lower temperatures. The normalized standard deviation (SD) of each distribution allows us to qualitatively distinguish three regimes of the $I_C$ switching process.[31, 32] Below $T = 0.7$ K, the SD is nearly temperature independent, indicating that the $I_C$ switching is governed by the MQT process.[21, 22] For 0.7 K < *T* < 1.1 K, the SD is proportional to temperature, which is due to the TA process.[23] Above 1.1 K, the SD decreases when temperature increases, and this can be explained by the PD process.[24, 25] A similar temperature dependence of the SD has already been observed in other nanohybrid JJs using graphene[31, 32] and semiconductor nanowires.[33, 40] We have reported the first experimental observations of the MQT behavior from TI-based JJs, and the observed MQT temperature, $T^*_{MQT}$, has the highest value among the nano-hybrid JJs. The physical origin of the highest $T^*_{MQT}$ will be discussed later.

For a quantitative analysis of the $I_C$ distribution, we define the switching probability[23] $P(I_C) = [\Gamma(I_C)/(dI/dt)]\left\{1 - \int_0^{I_C} P(I)dI\right\}$, where $\Gamma(I_C)$ is the escape rate, which depends on the escape regimes, and $dI/dt$ is the current sweep rate. Representative data (symbols) of the $I_C$ distribution and corresponding $\Gamma(I_C)$ at different temperatures are displayed in Figure 4a-b, overlaid by theoretical curve fits. For the data obtained at $T = 0.3$ K, we used the escape rate belonging to the MQT regime (see the Methods). The best fit (solid line) was obtained for parameters such as the fluctuation-free switching current $I_{C0} = 30.0$ µA and the junction capacitance $C = 29$ fF, which are similar to those obtained from the graphene-based JJ[31] in the MQT regime. The capacitance $C$ may be caused by a parasitic capacitance[28] formed between the source and drain electrodes via the conducting Si substrate or diffusive electronic transport[41] in the NW.

When the temperature is increased, the thermally activated escape of the phase particle from the local minima of the washboard potential becomes more dominant than the MQT process. The $P(I_C)$ and $\Gamma(I_C)$ data obtained at $T = 1.1$ K were fitted using the escape rate in the TA regime[23] (see the Methods) with the parameters of $I_{C0}$, $C$, and the escape temperature $T_{esc}$. Figure 4c depicts $T_{esc}$ over the entire temperature range in this experiment. The saturation of $T_{esc}$ below $T = 0.8$ K indicates that the MQT is the governing $I_C$ switching mechanism in that temperature region, as opposed to the TA. Here, the crossover temperature between the MQT and the TA regimes are determined to be $T^*_{MQT} = 0.8$ K. Because the shapes of the $P(I_C)$ distributions are almost identical in the MQT regime, nearly constant SD values are also expected, as already seen in Figure 3c.

For 0.8 K $< T <$ 1.1 K, $T_{esc}$ and the normalized SD increase linearly with the bath temperature $T$, which is consistent with the TA model. The close overlap between $T_{esc}$ and $T$, as depicted by the dashed line in Figure 4c, confirms the validity of the model. However, above 1.1 K, $T_{esc}$ estimated by the TA model decreases monotonically, deviating from the bath temperature. This discrepancy may be attributed to the PD process, where the thermally activated phase particle is retrapped in the neighboring potential well owing to a strong dissipation during the escape.[24, 25] Thus, the escape rate is suppressed in the PD regime, resulting in a very sharp $P(I_C)$ distribution and narrow SD at higher temperatures. The TA (dashed line) and PD (solid line) models were fitted to the $\Gamma(I_C)$ data taken at $T = 1.5$ K in Figure 4b, supporting the PD model as an appropriate switching mechanism (see the Methods). The results obtained after fitting (solid lines) other $P(I_C)$ data (symbols) at different temperatures are displayed in Fig. 3b, while the respective fitting parameters are plotted in Figure S4.

We now discuss the physical origin of the highly enhanced crossover temperature between the MQT and the TA regimes. The crossover temperature is theoretically given[48] by $T^*_{MQT} = a_t \hbar f_p / k_B$, where $a_t$ is a damping-dependent factor, $f_p$ is the Josephson plasma frequency, and $k_B$ is the Boltzmann constant. Because $f_p$ is proportional to $I_{C0}^{0.5}$ (see the Methods for detailed information), our observation of the highest $T^*_{MQT}$ among the nanohybrid superconducting junctions is attributed to the formation of the strongest Josephson coupling through the $\beta$-Ag$_2$Se TI NW. More quantitatively, $T^*_{\text{MQT}}$ is numerically obtained as $T^*_{\text{MQT,calc}} = 0.90$ K with $a_t = 0.76$ and $f_p = 155$ GHz for device D1, which is very close to the fitting result of the $P(I_C)$ data, $T^*_{\text{MQT}} = 0.8$ K. Our observation of the highest crossover temperature for sustaining the MQT behavior in the nanohybrid JJ would be useful to exploit a gate-tunable NW qubit[34, 35] combined with topological superconductivity.[5, 49]

In conclusion, we fabricated superconducting JJs using $\beta$-Ag$_2$Se TI NWs. The observed critical current reached the highest value of the nanohybrid JJs made of semiconductor NWs. The strong Josephson coupling strength enables us to observe the macroscopic quantum tunneling behavior, even at $T = 0.8$ K, which is the highest crossover temperature recorded to date. The measurement and analysis of the switching-current distribution reveals the underlying dynamics of the Josephson phase particle in β-Ag$_2$Se NW-based JJs. Our observations could contribute to development of superconducting qubits made of TI NWs.

**METHODS**

**$I_C(T)$ calculation:** Using the short and diffusive junction model[44], $I_C$ is given by the maximum value of the supercurrent expressed by

$$I_S(\varphi, T) = \frac{2\pi k_B T}{e R_N} \sum_{\omega > 0} \frac{2\Delta(T) \cos(\varphi/2)}{\delta} \tan^{-1} \frac{\Delta(T) \sin(\varphi/2)}{\delta}$$

and $\delta^2 = \Delta(T)^2 \cos^2(\varphi/2) + (\hbar\omega)^2$, where $R_N$ is the normal-state resistance of the junction, $\Delta$ is the superconducting gap, $\varphi$ is the phase difference between two superconducting electrodes, and $\omega$ is the Matsubara frequency that satisfies $\hbar\omega = \pi k_B T(2n+1)$ with the reduced Planck constant $\hbar$ and an integer $n$. The best-fitting line in Figure 2b was obtained using $R_N = 18.5$ Ω as a fitting parameter.

$I_C(B)$ **calculation:** When the width of the junction (diameter of the NW), $w$, is less than the magnetic length, $\xi_B = (\Phi_0/B_0)^{1/2}$, the magnetic-field dependence of $I_C$ can be expressed as[50]

$$eR_N I_C = \frac{4\pi k_B T}{r} \sum_{n=0}^{\infty} \frac{\Delta^2/(\Delta^2+\omega_n^2)}{\sqrt{2\left(\frac{\omega_n+\Gamma_B}{E_{Th}}\right)} \sinh\sqrt{2\left(\frac{\omega_n+\Gamma_B}{E_{Th}}\right)}},$$

where $\Gamma_B = De^2B^2w^2/6\hbar$ is the magnetic depairing energy, $D$ is a diffusion coefficient, $\omega_n = \pi k_B T(2n+1)$ is the $n$-th Matsubara energy with an integer $n$, $E_{Th}$ is the Thouless energy, and $r = R_B/R_N$ with the barrier resistance $R_B$. The solid line in Figure 2d is the best-fit result obtained with the parameter $r = 0.02$.

**Escape rate calculations:** In the MQT regime[22], the escape rate is given by $\Gamma_{MQT} = 12\omega_p(3\Delta U/\hbar\omega_p)^{1/2}\exp[-7.2(1+0.87/Q)\Delta U/\hbar\omega_p]$, where $\omega_p = \omega_{p0}(1-\gamma^2)^{1/4}$ is the Josephson plasma frequency, $\omega_{p0} = (2eI_{C0}/\hbar C)^{1/2}$ is the plasma frequency at zero bias current, $C$ is the junction capacitance, $\gamma = I/I_{C0}$ the normalized current, $I_{C0}$ is the fluctuation-free switching current, $\Delta U = 2E_{J0}[(1-\gamma^2)^{1/2}-\gamma\cos^{-1}\gamma]$ is the barrier height of the tilted washboard potential (see the inset of Fig. 3b), $E_{J0} = \hbar I_{C0}/2e$ is the Josephson coupling energy, and $Q = 4I_C/\pi I_R$ is the quality factor.

The escape rate in the TA regime[23] is given by $\Gamma_{TA} = a_t(\omega_p/2\pi)\exp[-\Delta U/k_B T]$, where $a_t = (1+1/4Q^2)^{1/2}-1/2Q$ is a damping-dependent factor. In the PD regime,[24] the escape rate is given by $\Gamma_{PD} = \Gamma_{TA}(1-P_{RT})\ln(1-P_{RT})^{-1}/P_{RT}$, where $P_{RT}$ is a retrapping probability. $P_{RT}$ is obtained by integrating the retrapping rate $\Gamma_{RT} = \omega_{p0}[(I-I_{R0})/I_{C0}](E_{J0}/2\pi k_B T)^{1/2}\exp(-\Delta U_{RT}/k_B T)$, where $I_{R0}$ is the fluctuation-free retrapping current, $\Delta U_{RT} = (E_{J0}Q_0^2/2)[(I-I_{R0})/I_{C0}]^2$ is the retrapping barrier, and $Q_0 = 4I_{C0}/\pi I_{R0}$ is the fluctuation-free quality factor.

## Figure Captions

**Figure 1.** (a) SEM image of $\beta$-Ag$_2$Se NWs grown on $c$-Al$_2$O$_3$ substrate. The inset in (a) is a magnified image of the NW tip, where the scale bar is 200 nm. (b) TEM image of $\beta$-Ag$_2$Se NW. (c) High-resolution TEM image of (b). The inset in (c) is the fast-Fourier transform pattern indexed to the orthorhombic $\beta$-Ag$_2$Se structure along the [1$\bar{1}$0] zone axis. (d) SAED pattern of a $\beta$-Ag$_2$Se NW along the [1$\bar{1}$0] zone axis.

**Figure 2.** (a) Temperature dependence of current-voltage (*I-V*) characteristics. The bias current was swept from negative to positive values. $I_C$ and $I_R$ indicate the critical and return currents, respectively. Inset: SEM image of a typical $\beta$-Ag$_2$Se NW Josephson device. Scale bar is 1 µm. (b) Temperature dependence of $I_C$ (filled symbols) and $I_R$ (empty symbols). Dashed line is a best-fit result using short and diffusive junction model. (c) Color scale plot of *dV/dI* as a function of bias current and magnetic field. (d) Magnetic-field dependence of $I_C$ (empty symbols) at $T = 0.3$ K. Solid line is a theoretical fit using the narrow junction model.

**Figure 3.** (a) *I-V* curves (black solid lines) recorded repeatedly (4,000 times) and switching current histograms (red) obtained using threshold voltage $V_{th} = 30$ µV. (b) Switching current distributions obtained at different temperatures. The symbols indicate the measured experimental data, and the solid lines are the best-fit results along with MQT, TA, PD model, respectively. Inset: schematic diagram of the tilted-washboard potential and three different types of the escape processes of the Josephson phase particle. (c) Temperature dependence of the normalized standard deviation for each switching current distribution.

**Figure 4.** Detailed plot of (a) the switching current distribution and (b) escaping rate in the PD, TA, and MQT regimes, respectively. Here, the symbols refer to the experimental data, and the solid lines are theoretical fit using the PD, TA, and MQT models. In the PD regime, the dashed line indicates a close fit of the TA model. (c) The escape temperate ($T_{esc}$) vs. bath temperature ($T$) plot obtained from TA (circle) and PD (triangle) models, respectively. $T^*_{MQT}$ ($T^*_{TA}$) indicates the crossover temperature between MQT (TA) and TA (PD) regimes.

## ASSOCIATED CONTENT

**Supporting Information**.

This material is available free of charge via the Internet at http://pubs.acs.org.

## AUTHOR INFORMATION

**Corresponding Author**

[*]E-mail: (B.K.) bongsoo@kaist.ac.kr; (Y.J.D) yjdoh@gist.ac.kr

**Author Contributions**

The manuscript was written with contributions from all authors. All authors have given approval to the final version of the manuscript. [†]These authors contributed equally to this work.

**Notes**

The authors declare no competing financial interest.

**ACKNOWLEDGMENT**

We are grateful to Gil-Ho Lee for fruitful discussions. This work was supported by the NRF of Korea through the Basic Science Research Program (Grant No. 2015R1A2A2A01006833) and the SRC Center for Quantum Coherence in Condensed Matter (Grant No. 2016R1A5A1008184).

**Table of Contents Graphic**

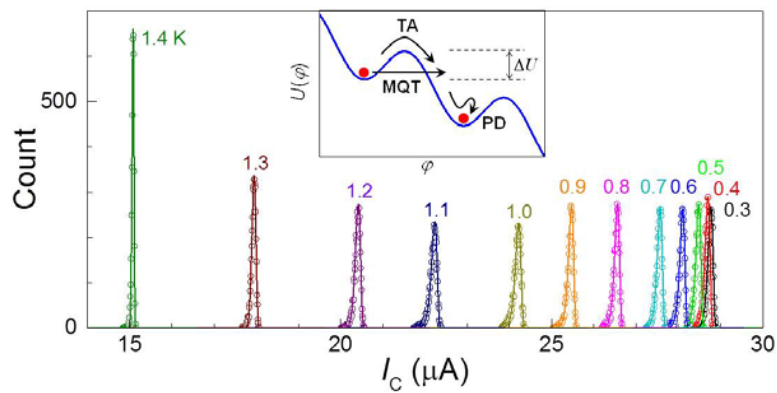

**Figure 1.**

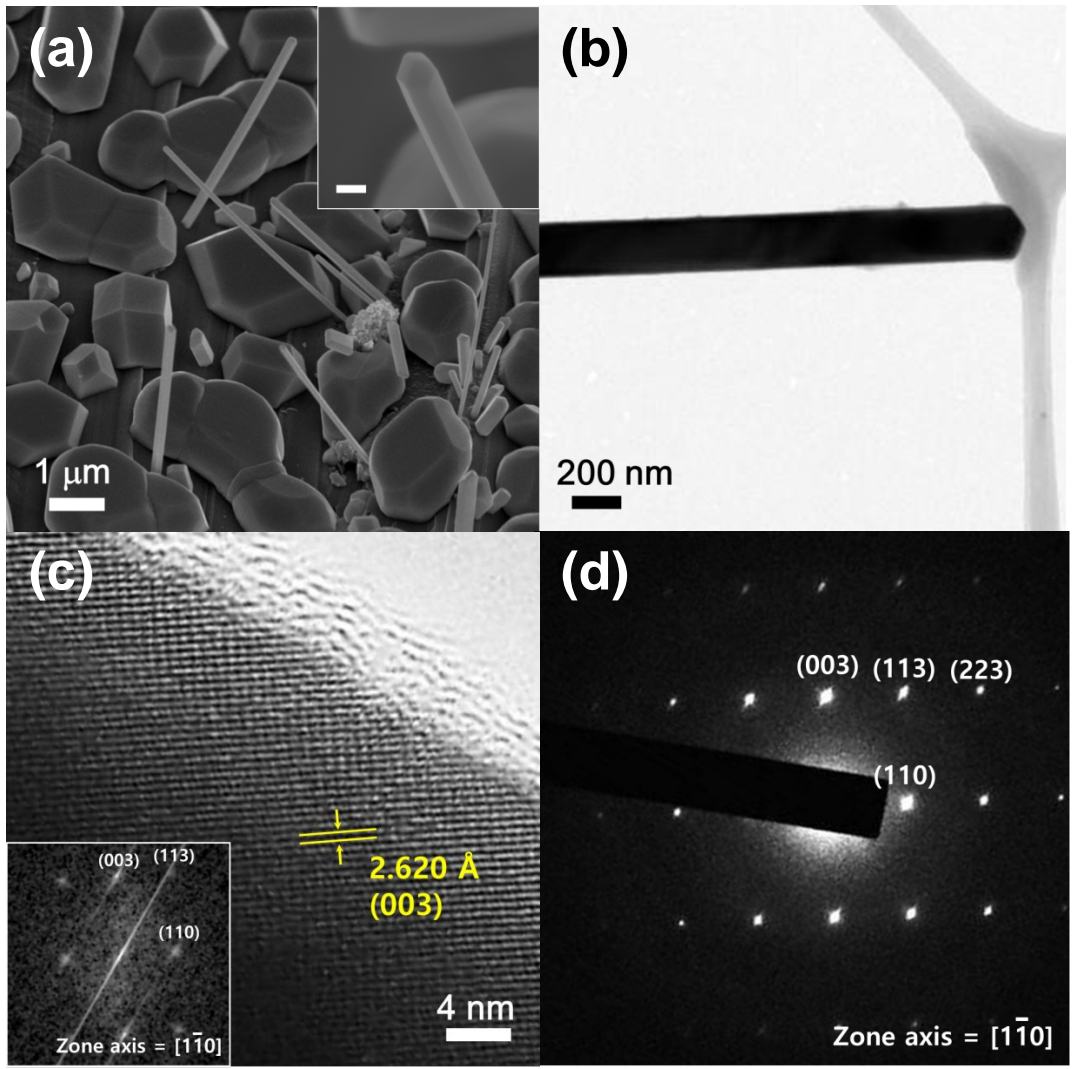

**Figure 2.**

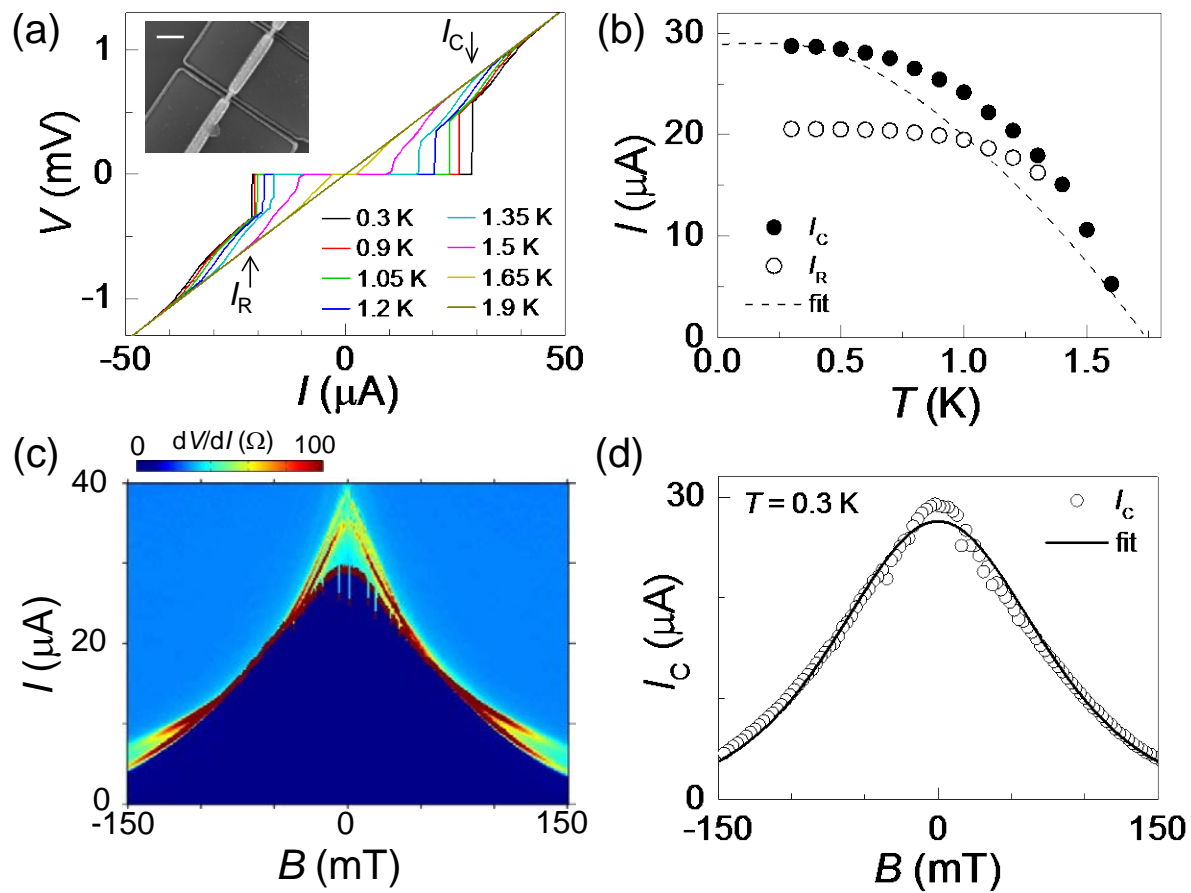

**Figure 3.**

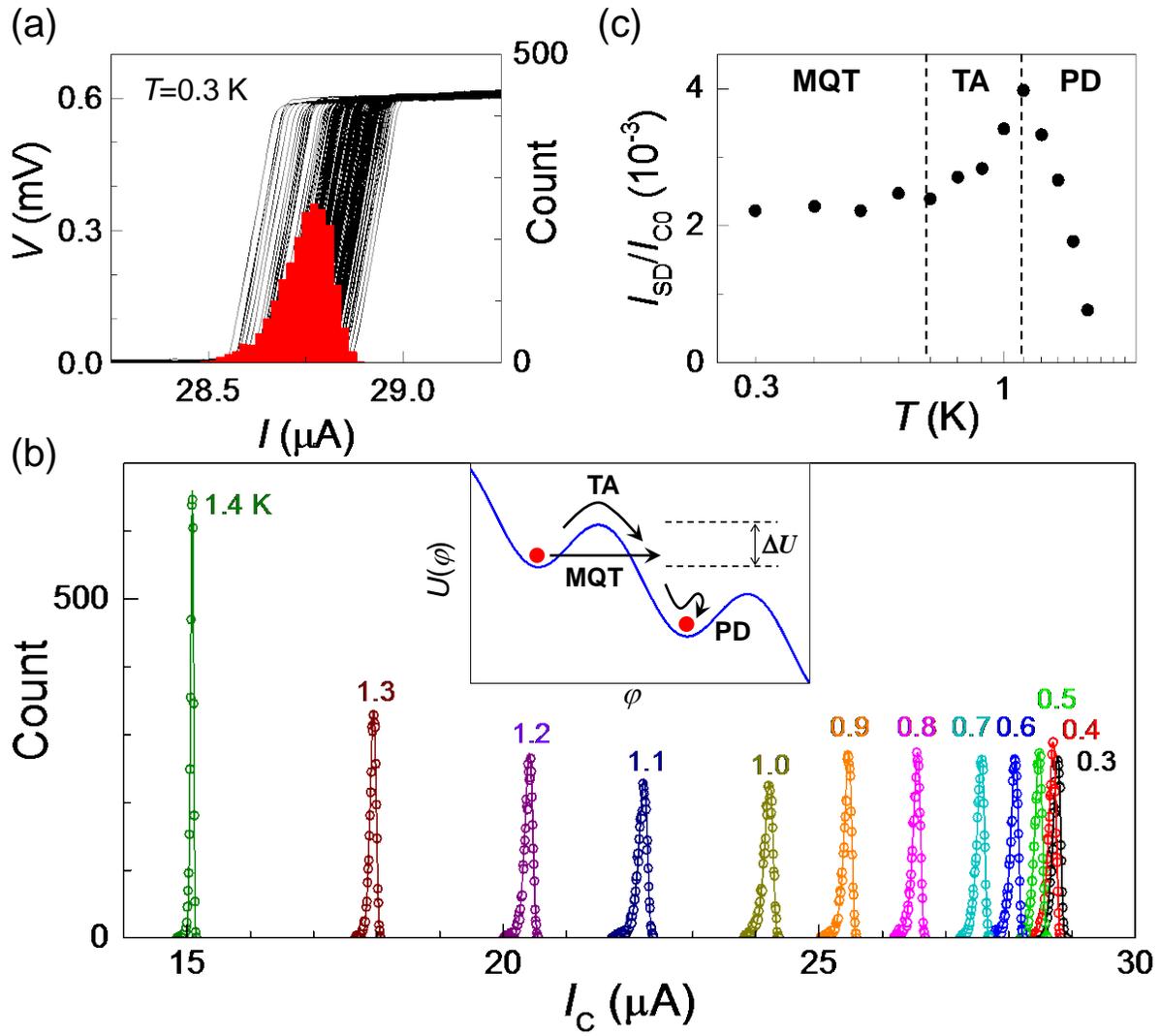

**Figure 4.**

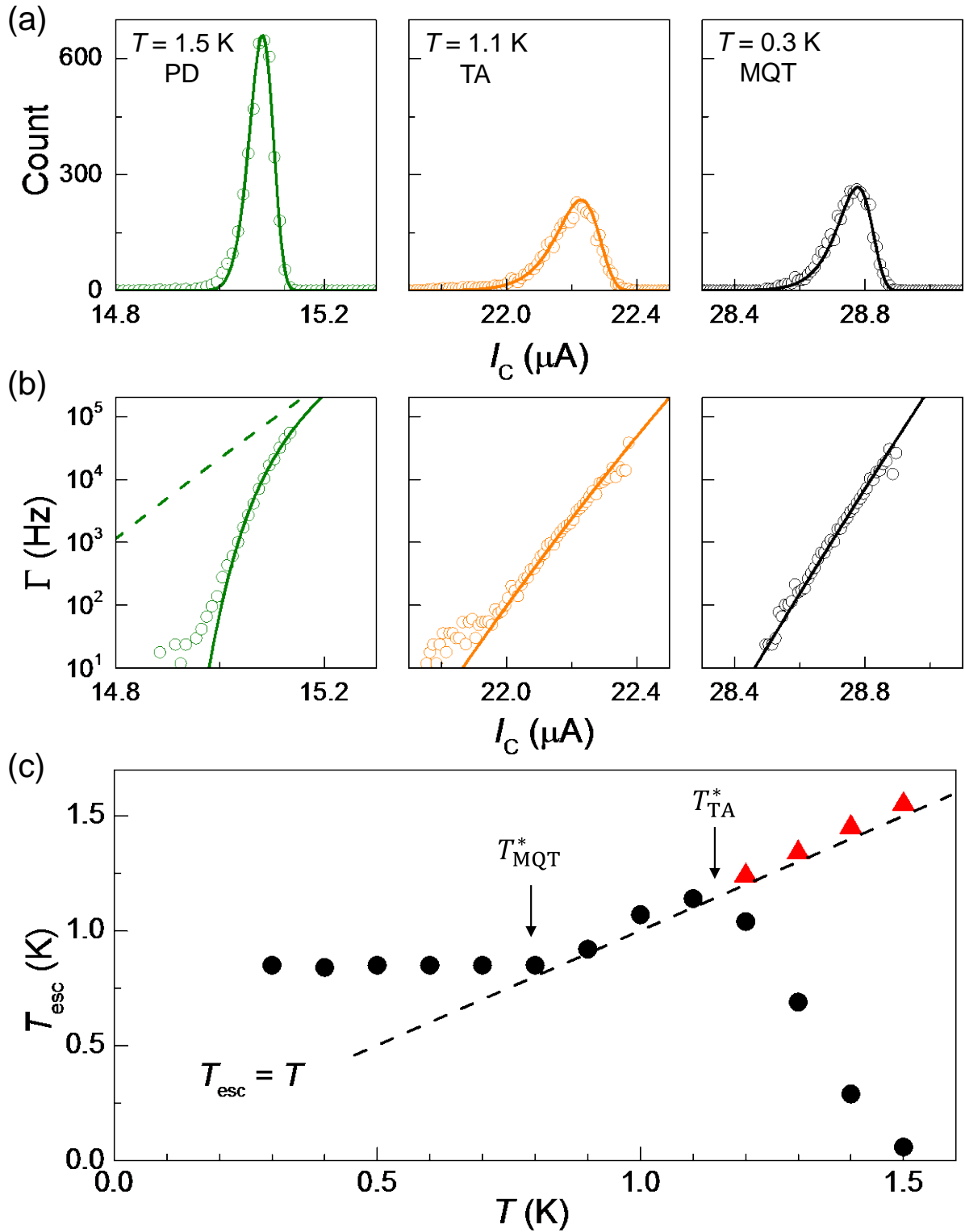

**Supplementary Information**

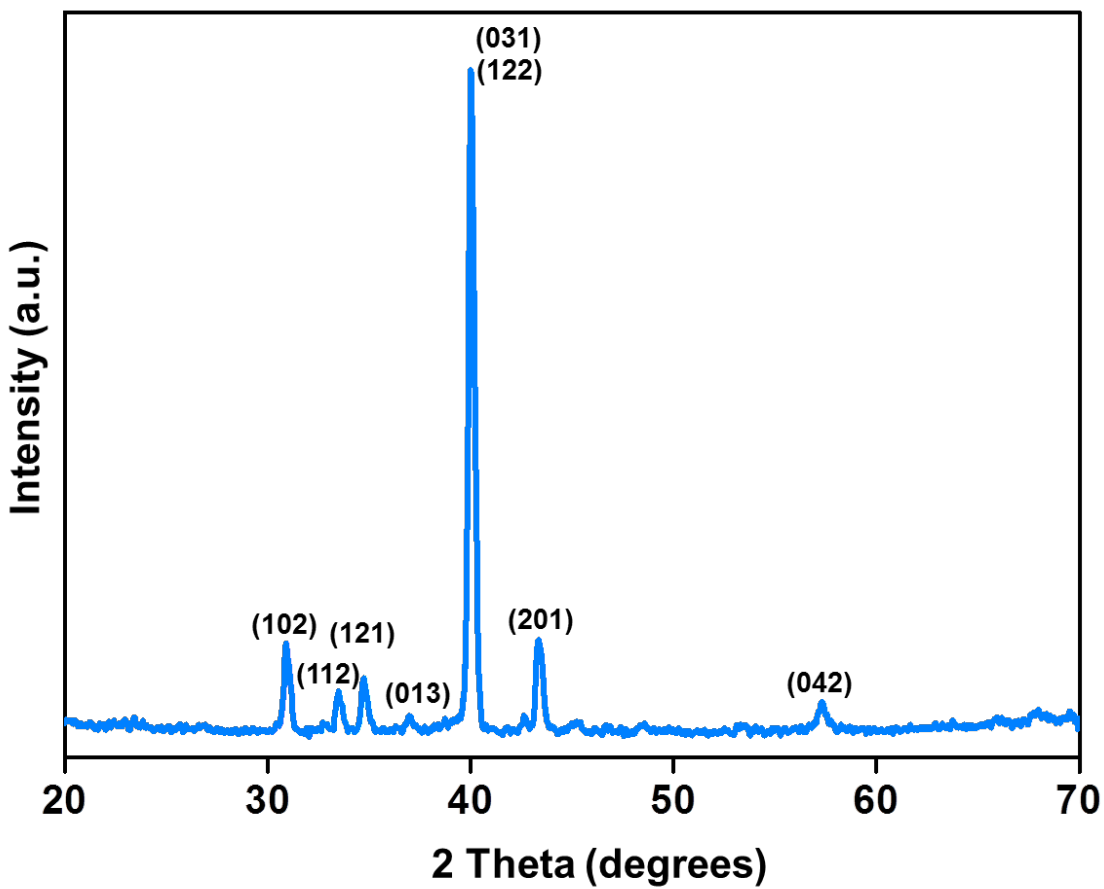

**Figure S1.** The X-ray diffraction pattern obtained from as-grown NWs on $c$-Al$_2$O$_3$ substrate. All the diffraction peaks are indexed to an orthorhombic β-Ag$_2$Se crystal structure (JCPDS card No. 01-071-2410).

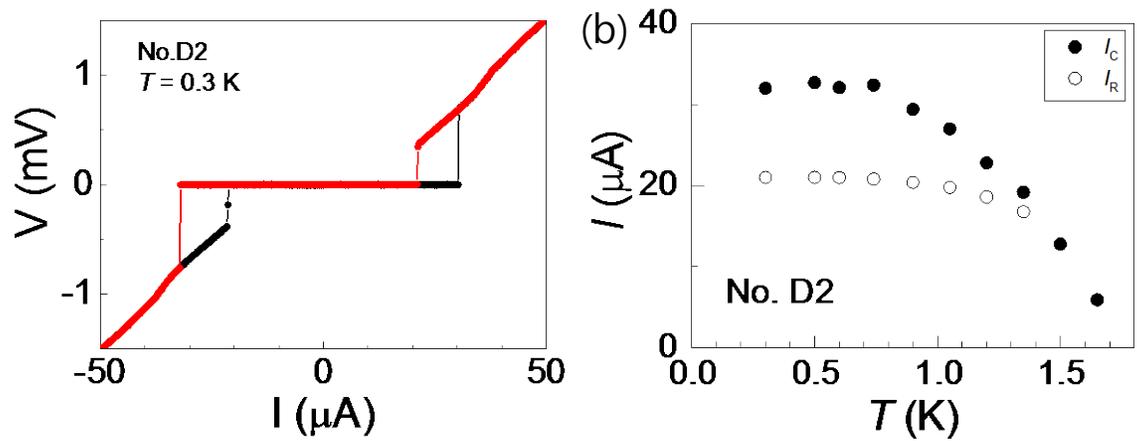

**Figure S2.** (a) *I-V* characteristic curve obtained from device **D2**. Black (red) line and symbols are from the forward (reverse) sweep of bias current. $I_C$ = 32.0 µA corresponds to $J_C = 1.4 \times 10^5$ A/cm$^2$. (b) Temperature dependence of $I_C$ and $I_R$.

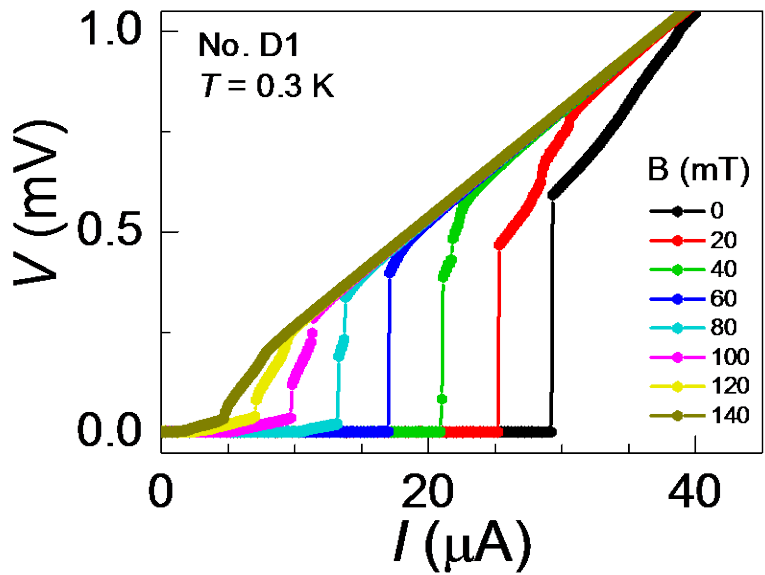

**Figure S3.** Forward swept *I-V* curves at different magnetic field.

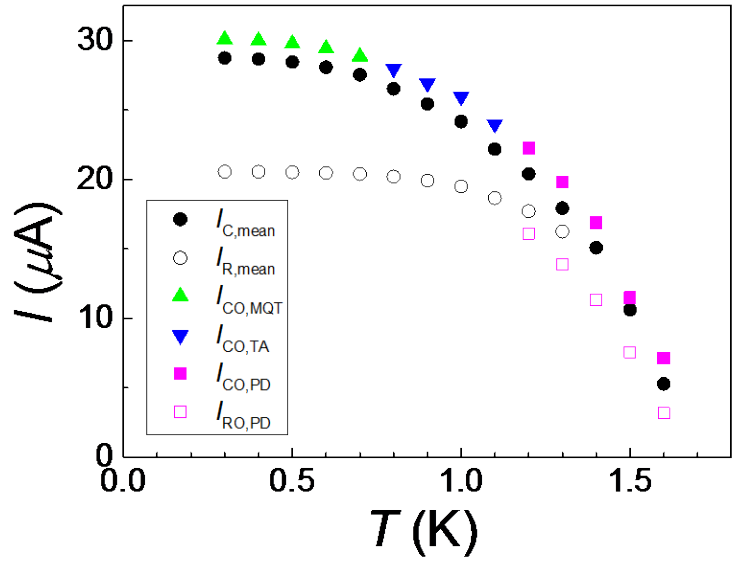

**Figure S4.** Temperature dependence of various fitting parameters used for the escape rate calculations. The fluctuation-free switching currents ($I_{C0}$) were obtained using MQT, TA and PD models, respectively. The fluctuation-free return current ($I_{R0}$) was obtained using PD model. $I_{C,mean}$ and $I_{R,mean}$ are the averaged experimental values of $I_C$ and $I_R$, respectively.